\documentstyle[11pt,epsf,wrapfig]{article}
\begin{document}
\pagestyle{empty}

\begin{flushleft}
\Large
{SAGA-HE-111-96  
\hfill October 27, 1996}  \\
\end{flushleft}
 
\vspace{2.5cm}
 
\begin{center}
 
\LARGE{{\bf Flavor Asymmetry $\bf \bar u-\bar d$ in the Nucleon}} \\
 
\vspace{1.2cm}
 
\LARGE
{S. Kumano $^*$ }         \\
 
\vspace{0.8cm}
  
\LARGE
{Department of Physics}         \\
 
\vspace{0.1cm}
 
\LARGE
{Saga University}      \\
 
\vspace{0.1cm}

\LARGE
{Saga 840, Japan} \\

\vspace{1.8cm}
 
\LARGE
{Talk given at the Circum-Pan-Pacific Workshop on} \\

\vspace{0.3cm}

{High Energy Spin Physics} \\

\vspace{0.7cm}

{Kobe, Japan, Oct. 2 -- 4, 1996 (talk on Oct. 2, 1996)}  \\
 
\end{center}
 
\vspace{1.5cm}

\vfill
 
\noindent
{\rule{6.cm}{0.1mm}} \\
 
\vspace{-0.2cm}
\normalsize
\noindent
{* Email: kumanos@cc.saga-u.ac.jp. 
   Information on his research is available}  \\

\vspace{-0.6cm}
\noindent
{at http://www.cc.saga-u.ac.jp/saga-u/riko/physics/quantum1/structure.html.} \\

\vspace{+0.5cm}
\hfill
{\large to be published in proceedings}

\vfill\eject
\setcounter{page}{1}
\pagestyle{plain}
\begin{center}
 
\Large
{Flavor asymmetry $\bar u-\bar d$ in the nucleon} \\
 
\vspace{0.5cm}
 
{S. Kumano $^*$}             \\
 
{Department of Physics, Saga University}      \\

{Honjo-1, Saga 840, Japan} \\

\vspace{0.7cm}

\normalsize
Abstract
\end{center}
\vspace{-0.30cm}

We give a brief summary of the Gottfried sum rule
and a flavor asymmetric distribution $\bar u-\bar d$ in the nucleon.
First, experimental history of the sum-rule studies is
discussed. Second, future prospects for studying the asymmetry
at Fermilab and RHIC are discussed. We also comment on
possible nuclear modification.

\vspace{0.6cm}


\noindent
{\bf 1. Introduction}

\vspace{0.2cm}

Unpolarized parton distributions in the nucleon are now well
understood by analyzing abundant high-energy experimental data.
However, light antiquark distributions had been considered
flavor symmetric until rather recently. It is because $u$ and $d$ quark
masses are very small compared with typical $\sqrt {Q^2}$ 
in deep inelastic processes. 
Therefore, equal mounts of $u\bar u$ and $d\bar d$
pairs are produced perturbatively through gluon splitting.
Although it was somewhat suggested in the SLAC data in 1970's,
the NMC conclusion of the $\bar u/\bar d$ asymmetry in 1991 was surprising.

As far as theory is concerned, there are several proposed models
for explaining the NMC result.
Those include Pauli-blocking mechanism, mesonic clouds, diquark
model, and so on.
These ideas are not discussed in this paper due to lack of space.
Interested reader may read a summary paper in Ref. [1].

In spite of the NMC suggestion, it is still not obvious whether 
the Gottfried sum rule is in fact violated
due to a possible small-$x$ contribution.
Therefore, independent experimental information is necessary
for finding an accurate $\bar u-\bar d$ distribution.
The NA51 Drell-Yan experiment in 1994 suggests a large flavor asymmetry,
which is consistent with the NMC finding.
Right now, the Drell-Yan experiment is in progress at the Fermilab,
and we expect to have much detailed results in the near future.
On the other hand, the other hadron collider RHIC should also be
able to provide information on the asymmetry.

In this paper, we first discuss experimental history
of the Gottfried sum rule and its relation to the flavor asymmetry
$\bar u-\bar d$ in section 2.
Then, Drell-Yan and W  production processes
are explained in sections 3 and 4 for finding the asymmetry
at Fermilab or RHIC.

\vspace{0.6cm}

\noindent
{\bf 2. Gottfried sum rule and flavor asymmetry $\bf \bar u-\bar d$}

\vspace{0.2cm}

The Gottfried sum rule is associated with the difference of proton
and neutron $F_2$ structure functions
measured in unpolarized electron or muon scattering.  
Because there is no fixed neutron target, the deuteron
is usually used for obtaining the neutron $F_2$ by subtracting
out the proton part. 
Using the parton-model expression:
$F_2(x,Q^2) = \sum_i e_i^2 x  [ q_i(x,Q^2) + \bar q_i(x,Q^2) ]$,
the baryon-number relation $\int dx (u_v-d_v)=1$, 
and isospin symmetry, we obtain
\begin{equation}
\int_0^1 \frac{dx}{x} \, [ F_2^p(x,Q^2) - F_2^n(x,Q^2) ]
 = \frac{1}{3} + \frac{2}{3} \int_0^1 dx [ \bar u(x,Q^2) - \bar d(x,Q^2) ]
\ \ .
\label{eqn:GINT}
\end{equation}
If the sea is flavor symmetric ($\bar u = \bar d$),
the second term vanishes and it becomes the Gottfried sum rule
\begin{equation}
\int_0^1 \frac{dx}{x} 
 [ F_2^p(x,Q^2) - F_2^n(x,Q^2) ] = \frac{1}{3} 
\ \ .
\label{eqn: GOTTFRIED}
\end{equation}
As it is obvious in the above derivation,
there is a serious assumption of the flavor symmetry in
light-antiquark distributions. Therefore, it is not a rigorous sum rule.
It is nevertheless interesting to test the sum rule because its violation
could suggest flavor asymmetric sea in the nucleon. 

\begin{wrapfigure}{r}{0.46\textwidth}
  \vspace{0.0cm}
\epsfxsize=6.0cm
\centering{\epsfbox{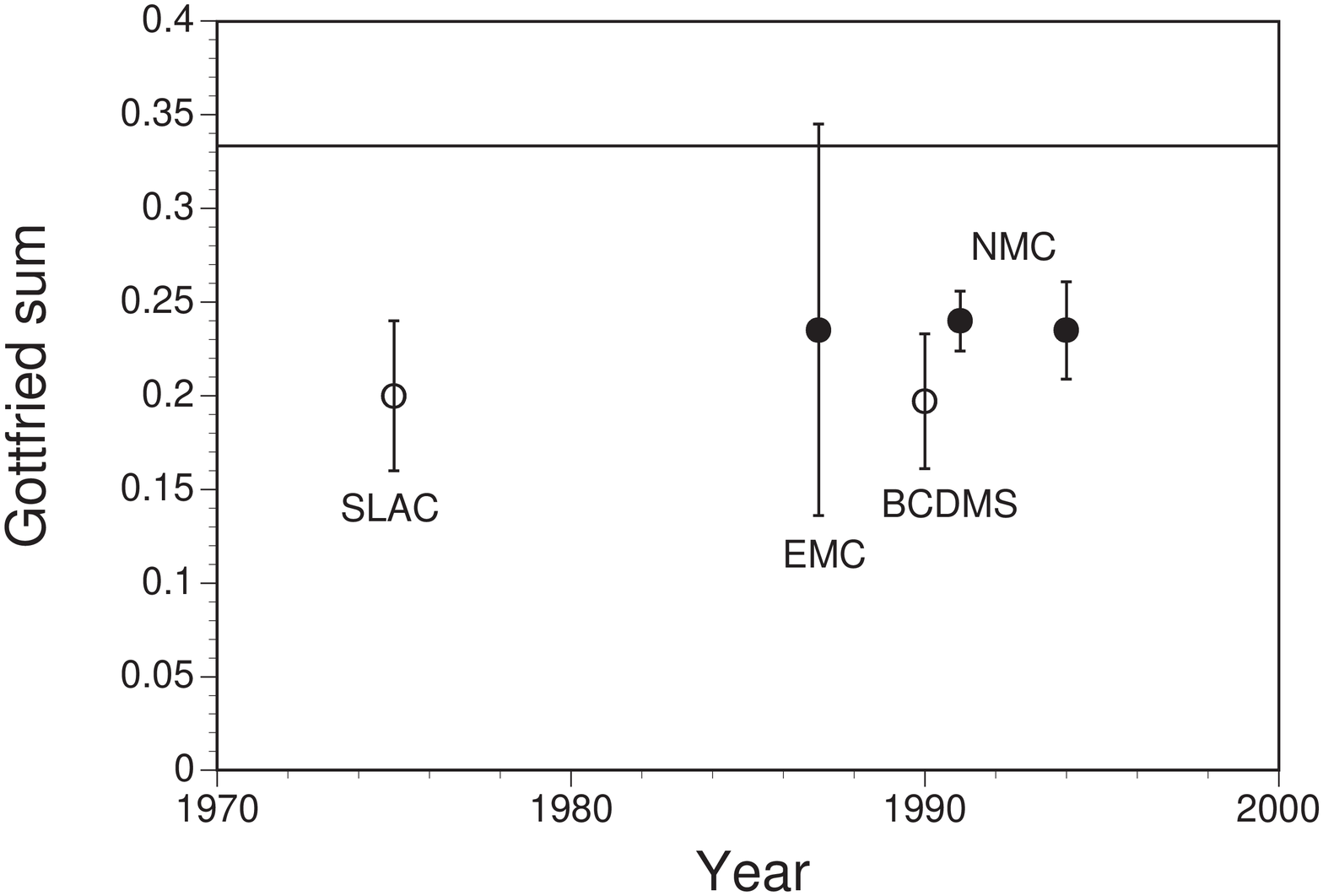}}
   \vspace{0.0cm}
{\footnotesize Fig.1 Experimental Gottfried sum.}
\end{wrapfigure}
The first test of the sum rule is studied at SLAC in the 1970's.
According to a SLAC paper in 1975, the integral is
$I_G(0.02,0.82) = 0.200 \pm 0.040$, where
we use the notation $I_G(x_{min},x_{max})$ for the Gottfried integral 
with minimum and maximum of the integral.
It is interesting to find a significantly smaller value
than the Gottfried sum 1/3.
It is, however, not conclusive enough to state that
the sum rule is violated due to a possible large contribution
from the smaller $x$ region. In Fig. 1, we show the SLAC result and
subsequent results by the EMC [$I_G(0,1) = 0.235  _{-0.099}^{+0.110}$], 
the BCDMS [$I_G(0.06,0.8) = 0.197 \pm 0.006(stat.) \pm 0.036(syst.)$],
and the NMC [$I_G(0,1) = 0.240 \pm 0.016$ in 1991 and
             $I_G(0,1) = 0.235 \pm 0.026$ in 1994].
The SLAC and BCDMS data are shown by open circles because small $x$
contribution is not estimated.
The NMC 1991 result is the first clear indication
of the sum-rule violation because of the small error.
From Eq. (\ref{eqn:GINT}), the NMC value $I_G$=0.235 could be explained
by the flavor asymmetry, namely a $\bar d$ excess over $\bar u$.
Next, we discuss other processes for testing the NMC $\bar u/\bar d$
asymmetry.

\vfill\eject

\noindent
{\bf 3. Drell-Yan process}

\vspace{0.2cm}

The NMC data are not enough for obtaining accurate $\bar u$ and
$\bar d$ distributions. As one of the other methods for
studying the flavor asymmetry, Drell-Yan process has been
studied. 
The Drell-Yan is a lepton-pair production process in hadron-hadron
collisions $A+B\rightarrow \ell^+\ell^- X$.
In the parton model, it is described by quark-antiquark
annihilation processes $q+\bar q\rightarrow \ell^+\ell^-$.
The cross section is given by
$\sigma_{_{DY}}\propto \sum_i e_i^2 \,
             [ q_i^A(x_1,Q^2) \bar q_i^B(x_2,Q^2)
              + \bar q_i^A(x_1,Q^2) q_i^B(x_2,Q^2) ]$,
where $Q^2$ is the dimuon mass squared $Q^2=m_{\mu\mu}^2$.
Drell-Yan p-n asymmetry at large $x_F$ is given by
\begin{equation}
A_{DY} \equiv \frac{\sigma^{pp}-\sigma^{pn}}{\sigma^{pp}+\sigma^{pn}}
      \approx \frac {[4u(x_1)-d(x_1)][\bar u(x_2)-\bar d(x_2)]}
                    {[4u(x_1)+d(x_1)][\bar u(x_2)+\bar d(x_2)]}
\ \ .
\end{equation}
It is directly proportional to the $\bar u-\bar d$
distribution. 

\begin{wrapfigure}{r}{0.43\textwidth}
  \vspace{0.0cm}
  \hspace{0.2cm}
\epsfxsize=4.5cm
\centering{\epsfbox{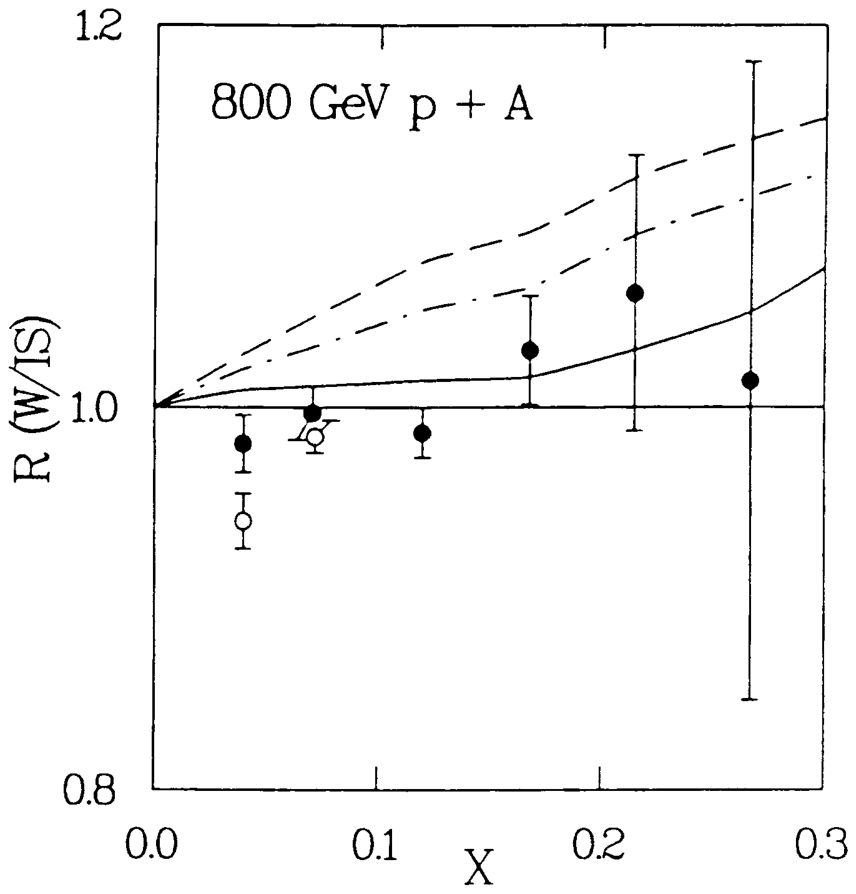}}\par 
   \vspace{0.2cm}
{\footnotesize Fig.2 Fermilab-E772 Drell-Yan data \par
\vspace{-0.1cm}\hspace{-0.5cm}
(taken from Ref. [2]).}
\end{wrapfigure}
There is a CERN-NA51 result:
$\bar u/\bar d = 0.51 \pm 0.04 (stat.) \pm 0.05 (syst.)$
at $x$=0.18, which is a clear indication of the $\bar u/\bar d$ asymmetry.
There are also Fermilab Drell-Yan data by E288 and E772;
however, the results are not conclusive enough for finding
the asymmetry. For example, the E772 data ($x_F>0.1$)
are shown in Fig. 2
together with theoretical curves of flavor asymmetry.
Because the ratio is given by
$\sigma_A/\sigma_{IS} \approx 1 
               + [(N-Z)/A] [\bar d(x)-\bar u(x)]/[\bar d(x)+\bar u(x)]$,
the deviation from unity indicates a $\bar u/\bar d$ asymmetry.
It is not obvious from the figure whether or not light antiquarks
are flavor symmetric.
However, the experimental analysis are currently in progress
at the Fermilab, so we expect have new accurate information in the near 
future.

\begin{wrapfigure}{r}{0.43\textwidth}
  \vspace{0.0cm}
\epsfxsize=6.0cm
\centering{\epsfbox{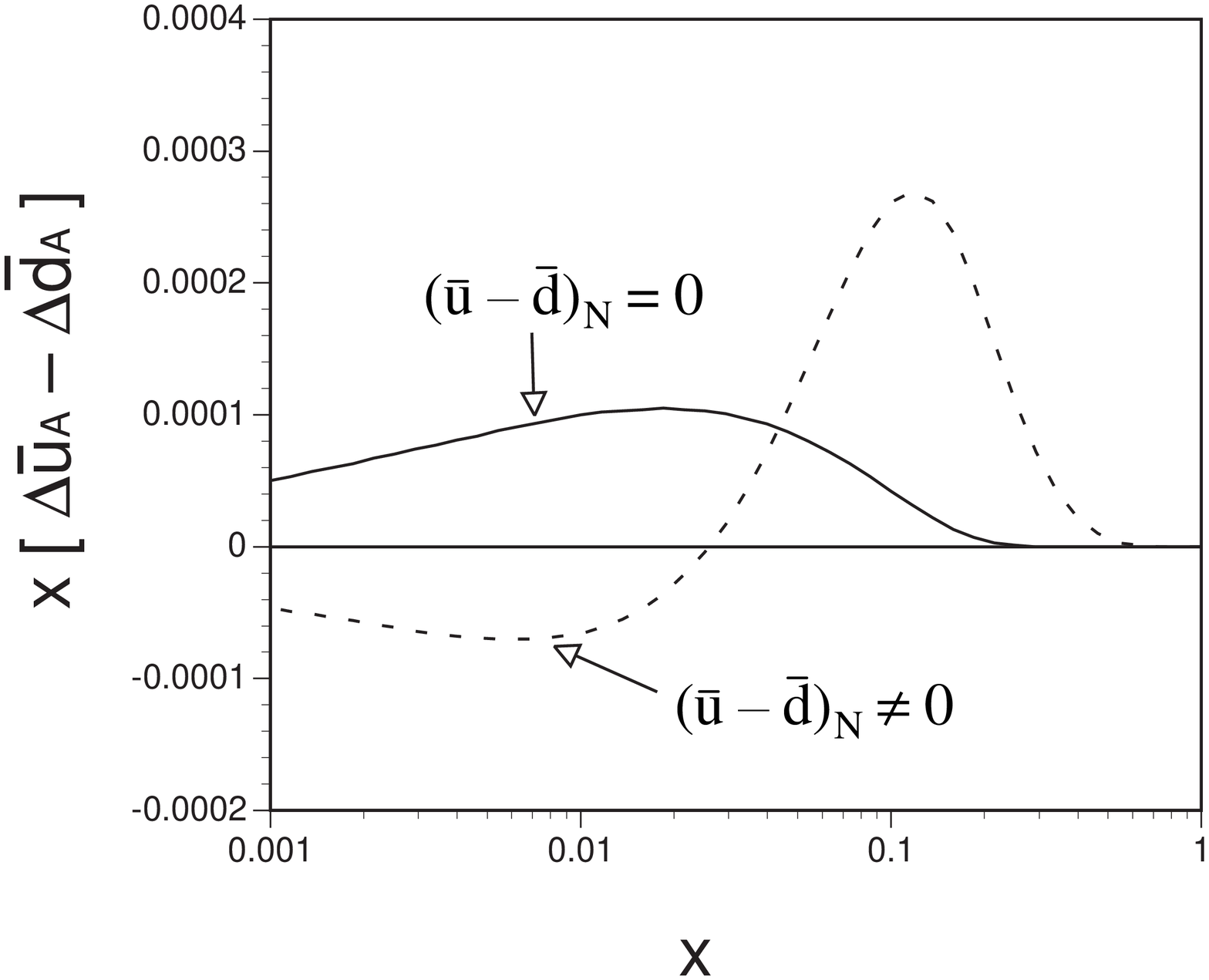}}
   \vspace{0.0cm}
{\footnotesize Fig.3 Nuclear modification of $\bar u-\bar d$ [3].}
\end{wrapfigure}
It should be noted that the E772 data are taken by nuclear targets.
The tungsten is a heavy nucleus with neutron excess.
Strictly speaking, they cannot be compared with the NMC data
due to possible nuclear modification.
Nuclear interactions may change the $\bar u-\bar d$ distribution. 
The modification is estimated in a parton-recombination model, which is one
of the ideas for explaining behavior of parton distributions
in the shadowing region.
Because of the neutron excess in the tungsten nucleus,
the $d\bar d$ recombination could occur more frequently than
the $u\bar u$. It means that a finite $\bar u/\bar d$ asymmetry
is produced even if it is symmetric ($\bar u=\bar d$) in the nucleon.
We show the $\bar u-\bar d$ distributions, which are created
by the recombinations in Fig. 3.
The detailed calculations are explained in Ref. [3].
Depending on the flavor distribution in the nucleon,
the results show interesting nuclear modification effects.
They are calculated at $Q^2$=4 GeV$^2$; however, the modification
should be larger at smaller $Q^2$ because of higher-twist nature.
Therefore, the nuclear effects are typically a few or several \%
in comparison with the $\bar u-\bar d$ distribution which is suggested
by the NMC data.
It is important to test the nuclear modification by the Drell-Yan
experiments for nuclear targets.

\vspace{0.6cm}

\noindent
{\bf 4. W production}
\vspace{0.2cm}

W production is investigated for measuring $\bar u-\bar d$ by several 
people [1]. Here, we discuss this topic based on Peng-Jansen studies [4].
The dominant subprocess of producing $W^+$ is $u+\bar d\rightarrow W^+$,
so that the $\bar d$ distribution could be extracted from $W^+$ production
data. On the other hand, the dominant process of $W^-$ production
is $d+\bar u\rightarrow W^-$, and $\bar u$ information can be obtained
instead of $\bar d$. This difference makes it possible to
find the asymmetric distribution $\bar u-\bar d$.
The $W^+$ production cross section is given by
\begin{eqnarray}
& & \frac{d \sigma_{p+p\rightarrow W^+}}{dx_F} 
      \propto \cos^2 \theta_c \,  
               [u(x_1) \bar d(x_2) + \bar d(x_1) u(x_2)] 
\nonumber \\
& & \ \ \ \ \ \ \ \ \ \ \ \ \ \ \ \ 
             + \sin^2 \theta_c \, 
               [u(x_1) \bar s(x_2) + \bar s(x_1) u(x_2)]  
\ \ .
\end{eqnarray}
Because the Cabbibo angle $\theta_c$ is small, the $\sin^2 \theta_c$ terms
are neglected for simplicity in the following discussions.
Then, the processes of producing $W^+$ are
$u(x_1)+\bar d(x_2)\rightarrow W^+$ and 
$u(x_2)+\bar d(x_1)\rightarrow W^+$.
The $W^-$ cross section is calculated in the similar way, and
we obtain the $W^\pm$ production ratio
\begin{equation}
R_{p+p}(x_F) \equiv 
\frac{d \sigma_{p+p\rightarrow W^+}/ dx_F}
     {d \sigma_{p+p\rightarrow W^-}/ dx_F } =
\frac {u(x_1) \bar d(x_2) +  \bar d(x_1) u(x_2)}
      {\bar u(x_1) d(x_2) +  d(x_1) \bar u(x_2)}
\ \ .
\end{equation}
At large $x_F$ (large $x_1$), the antiquark distribution
$\bar q(x_1)$ is very small, so that the above equation
becomes
$R_{p+p}(x_F \gg 0) \approx [u(x_1)/d(x_1)] [\bar d(x_2)/\bar u(x_2)]$,
which is directly proportional to the $\bar d/\bar u$ ratio.
On the other hand, we have
$R_{p+\bar p}(x_F \gg 0) \approx [u(x_1)/d(x_1)] [d(x_2)/u(x_2)]$
in the $p+\bar p$ reaction case. 
In the $x_F=0$ region, even though the $p+p$ ratio is still sensitive to the
flavor asymmetry: $R_{p+p}(x_F=0)=[u(x)/d(x)][\bar d(x)/\bar u(x)]$,
the $p+\bar p$ ratio is independent: $R_{p+\bar p}(x_F=0)=1$.
Therefore, the $p+\bar p$ reaction
is not a good way for finding the flavor asymmetry.
This situation is clearly shown in Fig. 4, where the $W^\pm$ production
ratio is calculated in the $p+p$ and $p+\bar p$ cases.

\vfill\eject
\begin{wrapfigure}{r}{0.46\textwidth}
  \vspace{0.0cm}
\epsfxsize=6.0cm
\centering{\epsfbox{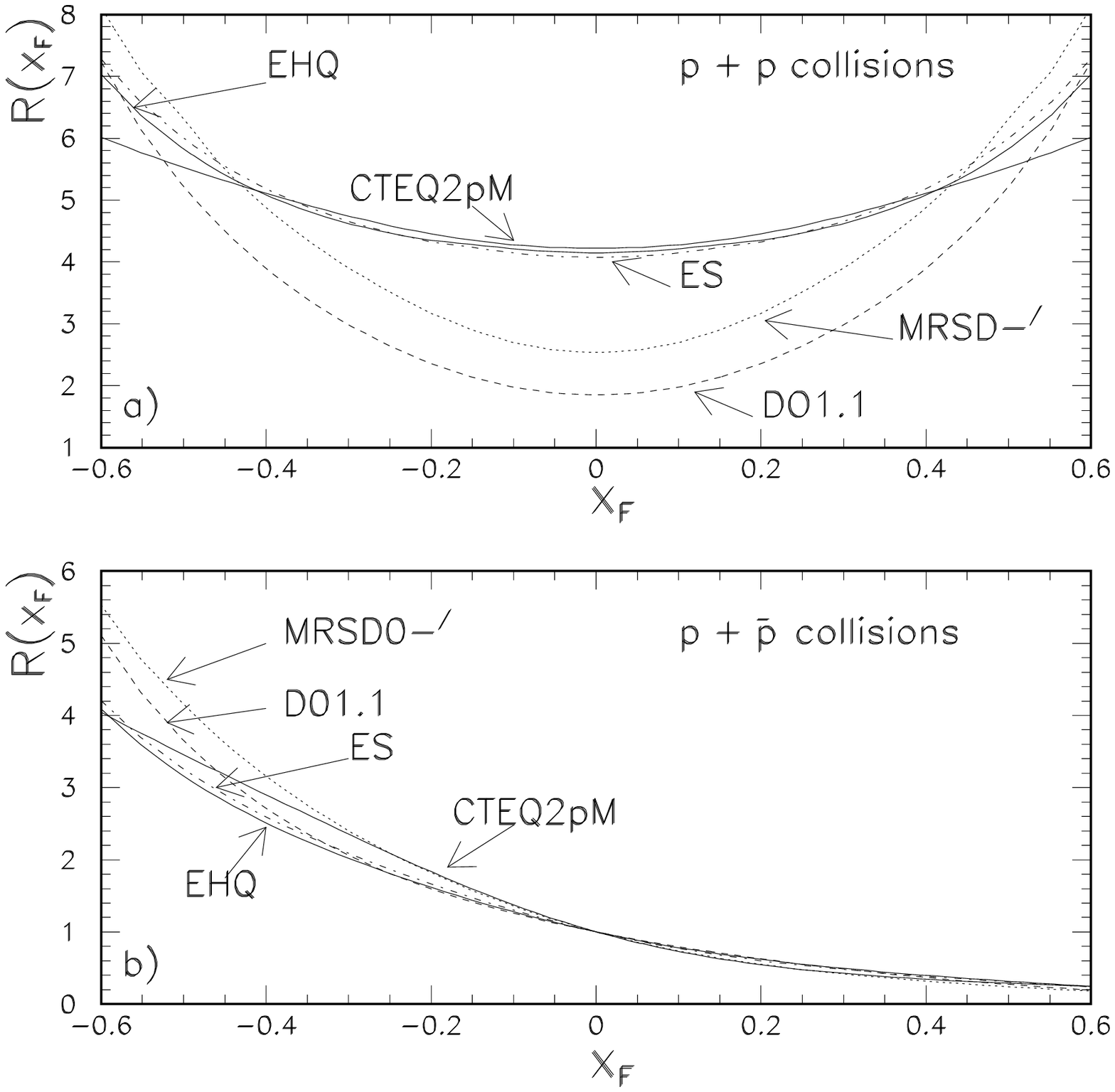}}\par 
   \vspace{0.3cm}
{\footnotesize Fig.4 W production in $p+p$ and $p+\bar p$\par
\vspace{-0.1cm}\hspace{+0.6cm}
reactions (taken from Ref. [4]).}
\end{wrapfigure}
The $W^\pm$ ratios in the $p+p$ and $p+\bar p$ reactions
are evaluated at $\sqrt s$=500 GeV by using various 
parametrizations for the parton distributions [4].
The distributions are evolved to the scale $Q^2=M_W^2$.
The figures a) and b) show the $p+p$ and $p+\bar p$ results
respectively. 
The dashed curve indicates results of using the flavor symmetric 
($\bar u=\bar d$) DO1.1 distributions. 
Others are the results for flavor asymmetric distributions 
(MRSD0$- '$, CTEQ2pM, ES, EHQ).
As we expected, the $p+p$ reaction is sensitive to 
the light antiquark flavor asymmetry
not only in the large $x_F$ region but also
in the $x_F\approx 0$ region.
On the other hand, the $p+\bar p$ reaction is almost insensitive 
to the asymmetry. The dependence appears only in the very small
$x_F$ region.

There are other possibilities such as Z and quarkonium production
processes for finding the $\bar u-\bar d$ distribution.
These topics as well as theoretical ideas are summarized in Ref. [1].
The flavor symmetry is an interesting topic for future investigations.
It is currently being studied at Fermilab and should be investigated
at RHIC. Furthermore, the asymmetry in polarized parton distributions
($\Delta\bar u/\Delta\bar d$) is an unexplored subject, which will
be studied for example at RHIC-SPIN.

\vspace{0.00cm}

\begin{center}
{\bf Acknowledgments} \\
\end{center}
\vspace{-0.17cm}

S. K. thanks RCNP for its financial support 
for participating in this conference.
This research was partly supported by the Grant-in-Aid for
Scientific Research from the Japanese Ministry of Education,
Science, and Culture under the contract number 06640406.

\vspace{0.2cm}

\noindent
{* Email: kumanos@cc.saga-u.ac.jp. 
   Information on his research is available}  \\

\vspace{-0.55cm}
\noindent
{\ \ \, at http://www.cc.saga-u.ac.jp/saga-u/riko/physics/quantum1/structure.html.} \\

\vspace{-0.30cm}

\begin{center}
{\bf References} \\
\end{center}
 
\vspace{-0.20cm}

\begin{description}{\leftmargin 0.0cm}

\vspace{-0.20cm}
\item{[1]}
S. Kumano, preprint SAGA-HE-97-96, to be submitted for publication.
\ \ \ \ \ \ 
We cannot list important papers in this reference due to lack of space.
The author hopes that the reader looks at the reference section of
this preprint.

\vspace{-0.20cm}
\item{[2]} P.L.McGaughey et al.(E772 collaboration),
                       Phys. Rev. Lett. 69 (1992) 1726.

\vspace{-0.20cm}
\item{[3]} S. Kumano, Phys. Lett. B 342 (1995) 339.

\vspace{-0.20cm}
\item{[4]} J. C. Peng and D. M. Jansen, Phys. Lett. B 354 (1995) 460.

\end{description}

\end{document}